\documentclass[conference]{IEEEtran}
\IEEEoverridecommandlockouts

\usepackage{amsmath,amsfonts,amssymb}
\usepackage{graphicx}
\usepackage{paralist}
\usepackage{color}
\usepackage{balance}
\usepackage{mathtools}
\usepackage{balance}
\usepackage{tabularx}
\usepackage{listings}
\usepackage{hhline}
\usepackage[linesnumbered,ruled]{algorithm2e}
\usepackage{bm}
\usepackage{url}
\usepackage{booktabs}
\usepackage[normalem]{ulem}

\usepackage{multirow} 
\usepackage{paralist}
\usepackage[inline]{enumitem}

\usepackage{array}
\newcolumntype{L}[1]{>{\raggedright\let\newline\\\arraybackslash\hspace{0pt}}m{#1}}
\newcolumntype{C}[1]{>{\centering\let\newline\\\arraybackslash\hspace{0pt}}m{#1}}
\newcolumntype{R}[1]{>{\raggedleft\let\newline\\\arraybackslash\hspace{0pt}}m{#1}}

\begin{document}
% ...


\iftrue 
\title{EDSC: An \underline{E}vent-\underline{D}riven \underline{S}mart \underline{C}ontract Platform}
\author{\IEEEauthorblockN{Mudabbir Kaleem\IEEEauthorrefmark{1}, Keshav Kasichainula\IEEEauthorrefmark{1}, Rabimba Karanjai\IEEEauthorrefmark{1}, Lei Xu\IEEEauthorrefmark{2}, Zhimin Gao\IEEEauthorrefmark{3}, Lin Chen\IEEEauthorrefmark{4}, Weidong Shi\IEEEauthorrefmark{1}}
\IEEEauthorblockA{\IEEEauthorrefmark{1}University Of Houston, USA\\
Email: \{mkaleem,kkasichainula,wshi3\}@uh.edu, rkaranja@cougarnet.uh.edu}
\IEEEauthorblockA{\IEEEauthorrefmark{2} University of Texas Rio Grande Valley ,USA\\
Email: xuleimath@gmail.com}
\IEEEauthorblockA{\IEEEauthorrefmark{3}Auburn University at Montgomery,USA \\ Email:mtion@msn.com}
\IEEEauthorblockA{\IEEEauthorrefmark{4}Texas Tech University,USA \\ Email: Lin.Chen@ttu.edu}}
\fi

\maketitle

\begin{abstract}
This paper presents EDSC, a novel smart contract platform design based on the event-driven execution model as opposed to the traditionally employed transaction-driven execution model. We reason that such a design is a better fit for many emerging smart contract applications and is better positioned to address the scalability and performance challenges plaguing the smart contract ecosystem. We propose EDSC's design under the Ethereum framework, and the design can be easily adapted for other existing smart contract platforms. We have conducted implementation using Ethereum client and experiments where performance modeling results show on average 2.2 to 4.6 times reduced total latency of event triggered smart contracts, which demonstrates its effectiveness for supporting contracts that demand timely execution based on events. In addition, we discuss example use cases to demonstrate the design's utility and comment on its potential security dynamics.


\end{abstract}

\begin{IEEEkeywords}
Smart contracts, Blockchain, Event-driven architecture, Scalability
\end{IEEEkeywords}
\section{Introduction}
\label{sec:introduction}
The advent of Bitcoin, in late 2008, demonstrated how a digital payment system could be implemented using a novel decentralized public data structure, i.e., a blockchain ~\cite{nakamoto2008bitcoin}. Additionally, Bitcoin also incorporated a built-in scripting framework that could be used for controlling the tokens, storing data, and specifying logic on the blockchain itself. This was a decentralized implementation of smart contracts ~\cite{szabo1997formalizing}. Smart contracts are essentially pieces of code that enforce the terms and procedures of an agreement or protocol digitally. However, Bitcoin's smart contract functionality was limited in its application, and it was not until the introduction of Ethereum ~\cite{buterin2014ethereum} that smart contracts took center stage in the cryptocurrency arena. Ethereum offered an integral Turing-complete programming language to create blockchain-based smart contracts, allowing for their employment in a wide range of potential use cases ~\cite{kehrli2016blockchain}.

Presently, smart contracts continue to grow in their utility and outreach. Since the launch of Ethereum, many alternative smart contract platforms have also emerged, which have gained considerable adaption and sizeable user bases ~\cite{io2019eos}~\cite{androulaki2018hyperledger}~\cite{lerner2019rootstock}~\cite{kiayias2017ouroboros}~\cite{mazieres2015stellar}. The majority of these platforms aim to overcome or readdress Ethereum's limitations and trade-offs, e.g., achieving higher throughput, decreasing computation costs, deploying a different consensus mechanism, etc. Although initially limited to token control and on-chain data access, smart contracts today are increasingly interfacing with real-world data and events, rapidly extending their application sphere. This interaction is enabled through oracles ~\cite{al2020trustworthy}~\cite{egberts2017oracle}, which are services designed to provide external information in the smart contract environment. To avoid a single point of compromise for such integration, many recent oracle projects ~\cite{chainlink}~\cite{provable}~\cite{augur} are adapting a decentralized approach for collecting and aggregating data.

Despite numerous innovations and advancements, the smart contract ecosystem's evolution has been stifled by various impediments, mostly prevailing in transaction performance (e.g., latency, throughput) and scalability domains. Although several projects ~\cite{al2017chainspace}~\cite{team2017zilliqa}~\cite{zamani2018rapidchain} have sought to address these concerns through imaginative solutions like sharding and execution parallelization; many complex design challenges remain unsolved for practical purposes. With the recent unprecedented growth of decentralized financial services (DeFi), an ever-increasing percentage of smart contracts are interfacing with oracle networks to fetch real-world information ~\cite{liu2020defioracles}. Since this interfacing is accomplished through two-way transactions on most platforms, the trend is bound to further burden an already congested system ~\cite{mcintosh2020ethereum}.

In the smart contract space, most existing popular platforms conform to the {\it transaction-driven execution model}. This means that all contract executions on these platforms are triggered by initiating transactions on the system through non-contract accounts. In this paper, we present an alternative smart contract platform design built on the event-driven execution model. The event-driven architecture pattern is a simple yet powerful distributed architecture pattern, proven to produce highly scalable and adaptable applications. The model enables communication by allowing participants to publish notifications of occurring events, along with subscribing to events of interest and being asynchronously notified of their occurrence by the system. The event-based methodology has previously been extensively studied in the context of systems and software engineering~\cite{eugster2003many}~\cite{parzyjegla2012engineering}~\cite{richards2015software}. We reason that a smart contract platform framework centered around the publish/subscribe paradigm will be a good fit for many emerging smart contract applications that demand or can benefit from timely execution. We also demonstrate that it will be, by design, better positioned to address the aforementioned issues hampering the ecosystem's progression.

This paper intends to describe an event-driven smart contract platform's architectural layout and implementation. We use the Ethereum architecture as the base template and outline the modifications required in its design to actualize our 
system. The rationale for this approach is that we assume most readers to be acquainted with Ethereum's mechanics, given that it is one of the most widely used smart contract platform to date. This familiarity, we hope, will allow the readers to draw parallels between the two models while enabling us to communicate our design 
succinctly. It is worth mentioning that although presented using Ethereum as the reference, the concept of event-driven smart contracts illustrated in this work can be extended to other smart contract platforms, consensus protocols, and programming models with trivial adjustments. The proposed design also has numerous advantages over the prior art that attempted to support events in the application layer. To the best of our knowledge, the proposed system is the first smart contract platform designed upon the event-driven execution model. We hope that our work will inspire further research in the direction of applying the event-driven communication paradigm to blockchains.

To summarize, the main contributions of the work and paper are:
\begin{inparaenum}[\bf(i)]
\item We propose an event-driven smart contract platform with native support for real-time event processing.
\item We provide the design of an event-based system using Ethereum as a reference target.
\item We describe the design's advantages in potential use cases and comment on its security aspects. 
\item We have performed an implementation using the Golang Ethereum client and conducted experiments where performance modeling results show on average a 2.2 to 4.6 times reduction in total latency of event triggered smart contracts, which demonstrates its effectiveness for supporting contracts that demand timely execution based on events.  
\end{inparaenum}

\section{Related Work}
\label{sec:background:related}


In recent years, the advancement of blockchains, smart contracts, and decentralized infrastructures have created an emerging frontier that combines traditional concepts of event-driven systems with blockchains. As such, there have been ongoing efforts to harness the benefits of this paradigm in blockchain-based smart contracts. Oracles and subscription-based payment models have sought to achieve this but have encountered limitations. Currently, oracle systems typically follow a pull-based model with the client contract requesting data from an off-chain source. Present designs favor off-chain implementations to incorporate the event-based subscription model and then interact with the blockchain.

We take as a case study the implementation of the IBM blockchain, which is built on Hyperledger \cite{cachin2016architecture}. Prior work by Hull \cite{hull2017blockchain} shows how event-based processing is used for data-centric applications. The commercial implementation and offering by IBM \cite{jennings} uses Java micro-services to listen for events form the blockchain using OpenLiberty. Blockchain provides the integrity of the process, whereas the java micro-service layer and OpenLiberty ensure it can have event-based transactions. However, apart from the implementation layer, it does not use the smart-contracts for any event-based transactions. Another commercial offering is provided by Amazon \cite{amazon}, which uses the Hyperledger Fabric and Ethereum as the underlying layer. Their implementation allows three distinct kinds of events to interact with the blockchain network, namely: 
\begin{inparaenum}[(i)]
    \item Block event, which occurs when a new block is added to the ledger;
    \item Transaction event and;
    \item Chaincode event, which can hold conditions for triggering events.
\end{inparaenum}
The triggering mechanism for these events relies on AWS Fargate to act as an event listener and then on Amazon Simple Queue Services to be processed by lambdas. Just like IBM, Amazon's implementation also relies on a layer of auxiliary services to enable event-driven architecture. 

Recent works remedying this limitation include EventWarden ~\cite{Li2020EventWardenAD}, where the authors propose a decentralized event-driven proxy that can interact with Ethereum-like blockchain networks and pass the transactions. This approach eliminates the use of auxiliary services similar to what IBM and Amazon were employing since this can be implemented directly onto the Ethereum network. It allows a user to create a proxy smart contract describing an event into the contract. Anyone in the blockchain network can trigger a release of the reserved transaction by calling the proxy contract and showing that the concerned event has been recorded into blockchain logs.

Another recent work Ethereum Alarm Clock ~\cite{merriam2015ethereum} allows a user to deploy a request contract with a future time limit on the Ethereum network. However, this project supports only one type of event, the arrival of a predefined time-frame. Work by Chao and Palanisamy ~\cite{Li2018DecentralizedPT} takes a similar approach to handle only events based on time.

In contrast, this paper tackles the fundamental limitations seen in ~\cite{jennings,amazon,merriam2015ethereum,Li2018DecentralizedPT} by proposing a smart contract platform based on the event-driven execution model, complete with a pub-sub scheme, which can be applied as a modification on the present Ethereum architecture or blockchains with smart contract support.

\section{Overview of EDSC} 
\label{sec:overview}
EDSC is built on the event-driven execution model using the publish/subscribe communication paradigm. In the publish/subscribe interaction scheme, components subscribe to events of interest, or to a pattern of events, and are subsequently asynchronously notified by the system when any event published matches their registered interest. In order to incorporate this paradigm into a smart contract platform, the platform design should provide the following basic features to the participating smart contracts and external accounts:
\begin{itemize}[leftmargin=*,itemindent=0.25cm]
\item \textbf{Event Definition:} Any external account or smart contract in the system is able to define/register new and unique event types in the system. This is analogous to defining a class in an object-oriented programming paradigm.
\item \textbf{Event Subscription:} Any smart contract in the system is able to subscribe or unsubscribe to a particular event type that is already defined in the system. At the time of subscription, the subscriber contract may specify additional logic that will be used by the system to evaluate whether to invoke it in response to the event of interest’s occurrence in the system.
\item \textbf{Event Publishing:} Any smart contract is able to publish an event that has already been defined in the system.
\end{itemize}
In order to provision the three fundamental features mentioned above, the smart contract system needs to incorporate the following functionality specific to the event-driven execution model:
\begin{itemize}[leftmargin=*,itemindent=0.25cm]
\item \textbf{Event Definition Maintenance:} The event templates are saved immutably in the system. This may be achieved in practice by referencing the event definitions on the blockchain itself, similar to how smart contract code is stored on-chain by reference in Ethereum.
\item \textbf{Subscription Information Maintenance:} The subscription information is also saved immutably in the system. This can also be achieved in practice by referencing the subscription information on the blockchain itself, similar to how smart contract code is stored on-chain by reference in Ethereum.
\item \textbf{Event Matching:} Every time a published event is processed, the system determines all the smart contracts which are subscribed to that particular event. The system also evaluates the corresponding subscription logics of all those subscriptions to determine which smart contracts to invoke in response to the publishing event.
\item \textbf{Event Queueing:} Based on the event matching, the system queues all the matching subscribed smart contracts for execution. Since the system is asynchronous, there are no guarantees as to when the subscription triggers will be executed. The system guarantees the queueing of these executions.
\end{itemize}
Since the publish/subscribe method is an anonymous and indirect communication paradigm, the system decouples the communicating entities i.e., the smart contracts in space and execution flow:
\begin{itemize}[leftmargin=*,itemindent=0.25cm]
\item \textbf{Space Decoupling:} The publishing and the subscribing smart contracts do not need to know each other since they are not required to address each other for communication. Hence, the event publisher does not maintain a record of all the smart contracts which will be evoked in response to its event publication. Likewise, the subscriber may subscribe to events from multiple sources without specifying them individually.
\item \textbf{Time Decoupling:} There is no provision for the publisher or the subscriber to run within any time constraint. The subscriber execution can be queued for a later time window (depending on future events).
\item \textbf{Execution Flow Decoupling:} The inherently asynchronous communication decouples execution flow from inter-contract communication. A smart contract is not blocked when sending a notification to an external contract. The system can handle the subscriber execution in response to the notification by running it concurrently or queueing it for later. The subscriber and publisher of events do not have to be synchronized in their execution.
\end{itemize}

\section{Advantages of EDSC} 
\label{sec:advantages}

Based on our basic design framework from Section ~\ref{sec:overview}, the proposed smart contract system will offer attractive advantages to the ever-evolving ecosystem of smart contracts:
\begin{itemize}[leftmargin=*,itemindent=0.5cm]
\item \textbf{Lower Fee for All:} We reason that the proposed platform will result in a majority of the system participants having to pay a lower gas fee for their executions, especially in a system that is highly interfaced with external oracle systems through oracle contracts. User smart contracts only pay for the gas for executing themselves. In other words, the transaction cost, which is now the cost of putting the event on-chain, is shared by all the subscribers collectively.
\item \textbf{Improved Security:} Ethereum smart contracts developed in Solidity have been marred with security issues centered around reentrancy and unexpected reverts~\cite{chen2020survey}. This is because, by design, an Ethereum transaction has to complete the contract execution in the current as well as called contracts before the transaction is considered complete. On the other hand, the event-driven paradigm is free from such vulnerabilities, since events are asynchronously published without waiting for the subscriber contracts to run. This offers better security guarantees.
\item \textbf{Less Network Clogging:} Having multiple smart contracts subscribe to a single event translates to lesser network usage as opposed to smart contracts requiring transactions to be broadcasted every time they need to execute or interface with an oracle provider. Also, only a single event needs to be recorded on-chain as opposed to multiple transactions. This is beneficial for freeing up vital network bandwidth, which has recently been clogged ~\cite{mcintosh2020ethereum}.
\item \textbf{Better Scalability:} We also claim that an event-driven system based on the proposed basic design is better positioned for employing parallel-processing and sharding solutions for scalability. This is because all executions are restricted in context to the currently executing contract, and event-based subscription triggering occurs asynchronously. All events can be posted to a shared global non-sharded trie for inter-shard communication borrowing from a similar concept in the Zilliqa project~\cite{team2017zilliqa}.
\end{itemize}

\section{EDSC System Design}
\label{sec:detailed:design}

We present the detailed design of the event-driven smart contract system based on the current Ethereum design framework. 
The rationale for this approach is that Ethereum is arguably the simplest and unarguably the most widely adopted smart platform to date. 
Using Ethereum’s design as the base reference will allow the readers to grasp the design proposals clearly and draw parallels between the two approaches. 
Note that the design presented is general and independent of any specific platform.

\subsection{Event Definition Trie}
All events have to be defined in the system before smart contracts can subscribe to or publish them. The event definitions must be stored in the system immutably and free from loss. In the Ethereum context, this may be achieved by requiring all nodes to maintain the global event definition data locally. This event definition data can then be referenced on the blockchain for immutability. This is analogous to how the current Ethereum design maintains the state of the system. In other words, the event definition trie will need to be added to the Ethereum system, as illustrated in \figurename~\ref{fig:eventtries}.

As mentioned in Section ~\ref{sec:overview}, any smart contract or external account in the system has the ability to define a new event type. This can be done by posting a special type of event that is already predefined in the system. The special event’s payload consists of the definition of the new event’s template. Any node of the network, when processing this event, adds the event definition to their local event definition database. An event definition consists of the attributes listed in \tablename ~\ref{tab:eventdef}.

\begin{table}
\centering
\caption{Event definition attributes.}
\label{tab:eventdef}
\scriptsize
\begin{tabular}{p{0.15\linewidth} | p{0.75\linewidth}}
\hline
\hline
 \textbf{Attributes}                              & \textbf{Description}   \\ 
\hline
\hline
\textbf{Event Identifier} &     This is a unique identifier for every event type that is defined in the system. One possible implementation is to use the hash of the entire event definition.     \\ \hline
\textbf{List of Variables} & A list of variables and their types which are posted as payload whenever this event is published in the network. \\ \hline
\textbf{Comments} & This part is for documentation purposes and can be used to write a description of the event, document the variables in the payload, for specifying other information for potential subscribers like event generation frequency or for any other information which the event definition originator desires. \\ 
\hline
\hline
\end{tabular}
\end{table}

\subsection{Event Subscription Trie}
Once an event has been defined in the network's subscription trie, the system participants, i.e., the smart contracts, can subscribe to such events. Subscribing to an event allows a smart contract to be executed in response to the particular event getting posted in the network. Whenever the subscribed event occurs, the subscriber smart contract's default callback function is asynchronously invoked, and the event's unique identifier and the payload are passed as arguments. The smart contract can then execute the desired functionality accordingly.

\begin{figure}
\begin{center}
\includegraphics[width=3.2in,angle=0]{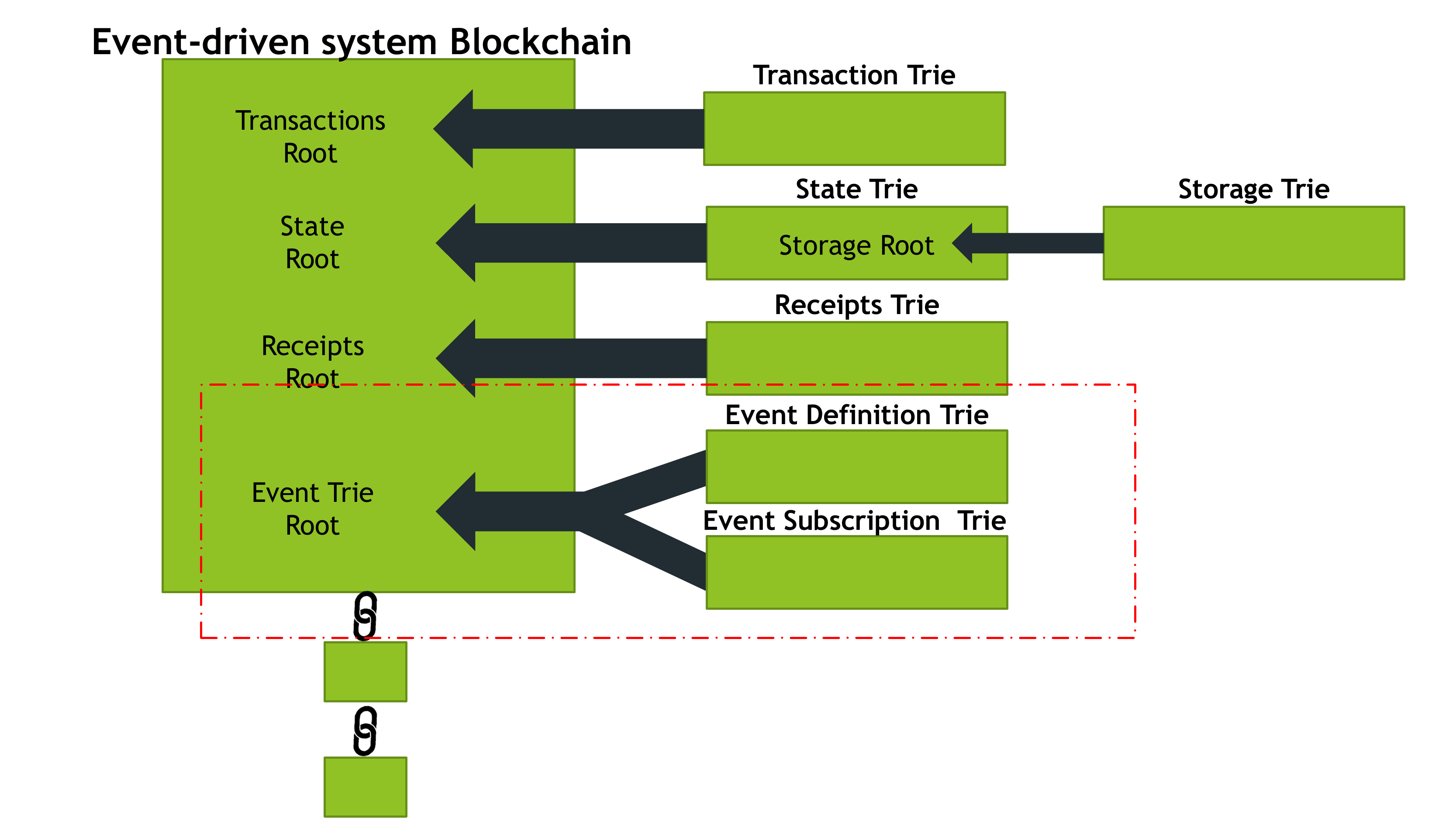}
\caption{Subscription Trie addition to Ethereum design.}
\label{fig:eventtries}
\end{center}
\end{figure}

Like event definitions, event subscription information also needs to be stored in the system immutably and without loss. We propose storing the subscription information in a similar manner to the event definition storage where it is stored across all nodes and referenced on the blockchain. In fact, the event definitions and subscription information can be combined into one trie which can be referenced on-chain, as shown in \figurename~\ref{fig:eventtries}.

Like event definition, event subscription also occurs through a special type of event that is predefined in the system. Smart contracts post this event to signal their desire to subscribe to a particular event type, which is passed as the payload of this predefined subscription event. The nodes of the network then update their subscription trie when this event is processed.
In addition to the event type, the system also allows subscribers to specify other parameters of their subscription which are summarized in in \tablename ~\ref{tab:eventsub}.

\begin{table}
\centering
\caption{Event subscription attributes.}
\label{tab:eventsub}
\scriptsize
\begin{tabular}{p{0.15\linewidth} | p{0.75\linewidth}}
\hline
\hline
 \textbf{Attributes}                              & \textbf{Description}   \\ 
\hline
\hline

\textbf{Event Identifier} &     The unique identifier for the event to subscibe to. \\ \hline
\textbf{Gas Rate} & Each subscriber contract at the time of subscription specifies the gas price that they will pay for their execution as a result of the subscription. The nodes decide which subscriptions to execute based on the gas price that they are willing to pay. So setting a higher gas price decreases the delay between event generation and subscription execution. Note that the design can be adapted to other public blockchain systems that apply alternative incentive mechanisms for running smart contracts. \\ \hline
\textbf{Subscription Fee} & The maximum fee that the subscriber is willing to pay the event publisher in order to be triggered by its generation (e.g., gas limit for event subscription in Ethereum). \\ \hline
\textbf{Publisher Identifier} & The subscriber also has the provision to only subscribe for execution when the event is published signed by a specific public key. This feature exists to allow a subscriber to only run in response to event publishers which they trust and to prevent spamming in the system. A subscriber may provide more than one public key. \\ \hline
\textbf{Block Rate} & The subscriber can also specify a block rate to run the subscription. For example, a subscriber may only want their subscription to be executed once every one hundred blocks for a frequently occurring event. \\ \hline
\textbf{Event Rate} & The subscriber can also specify the event rate to run the subscription. For example, a subscriber may only want their subscription to be executed once every hundredth instance of a specific event being generated. \\ \hline
\textbf{Subscription Logic (Constraints)} & The subscriber may also use the subscription logic field to specify any complex expression involving the block number, block time, event payload, event publisher public key, etc. This expression can then be evaluated to determine if the corresponding subscription should be triggered. The computational cost of evaluating the expression will be paid out of the subscriber’s account at a fixed system rate. \\
\hline
\hline
\end{tabular}
\end{table}

All the subscription parameters are stored in the subscription trie against each subscription and used by the system to determine which subscriptions to trigger in response to a generated event. Every time a smart contract makes a subscription to an event type, an entry against that event type is added in the subscription trie with the subscriber smart contract's address and all the subscription parameters provided.

\vspace{-5pt}

\subsection{Event Generation}
\label{sec:5.3}
Events can be generated in the system in three ways. Firstly, by external or non-contract accounts, which is similar to generating a transaction in the Ethereum system. An external account has to digitally sign such an event, similar to Ethereum transactions. Such events are generated outside the system and then broadcast in the network. Secondly, events can also be generated by smart contracts using a specific event generation opcode. Such events can be thought of as the analogous functionality for the CALL opcode in the Ethereum domain but working asynchronously. Thirdly, certain special events can be generated by the system itself as described in Section~\ref{sec:5.4}. The second and third types of events do not have to be digitally signed and only exist in the execution environment. Only the external account generated events are recorded on-chain, similar to Ethereum’s transactions. More on the topic follows under Section~\ref{sec:5.8}. 

An event generation message needs to contain the parameters summarized in \tablename ~\ref{tab:event}.

\begin{table}
\centering
\caption{Event message attributes.}
\label{tab:event}
\scriptsize
\begin{tabular}{p{0.15\linewidth} | p{0.75\linewidth}}
\hline
\hline
 \textbf{Attributes}                              & \textbf{Description}   \\ 
\hline
\hline
\textbf{Event Identifier} & The unique identifier of the event in the system. \\ \hline
\textbf{Publisher Identifier} & The public key of the event publisher is required when generating an event. The subscribers may use this information to subscribe to events from specific entities only. If the event is generated through an external account, then a digital signature corresponding to that public key is required to establish identity. \\ \hline
\textbf{Payload} & The event object also contains the event’s payload arguments as defined in the event definition. \\ \hline
\textbf{Subscription Fee} & This is the fee that any contract which subscribes to this event must pay to the event publisher when it runs in response to the event. This is different from the gas fee which is paid to the miners as the computation, network, and storage costs of running the smart contract. The subscription fee exists purely to incentivize event publication on the network. \\ \hline
\textbf{Inclusion Fee} & This is the fee that the publisher is willing to pay the miners for the inclusion of their generated event in the block. This is only required of external account generated events. \\ 
\hline
\hline
\end{tabular}
\end{table}

\subsection{Special Event Types}
\label{sec:5.4}
In addition to event definition and subscription/unsubscription events, there are two other special types of events in the system: the transaction event and the deploy event. A transaction event is an event to which every smart contract is subscribed by default and is triggered if the event contains that smart contract’s address in its payload. The transaction event is used to transfer tokens from one account/contract to another. The system automatically increments and decrements the receiver and sender’s balances depending upon the value specified in the payload when this event is processed. Since the transactions are also now events, the proposed system processes all transfer of tokens asynchronously too.

The deploy event is used to deploy a new smart contract in the system and specifies the contract code in its payload. It is analogous to using a transaction in Ethereum to deploy a new contract.

In addition to these two events, there can also be other special events that are system generated. Currently, we propose to have one special system generated event, which is the new block event. This event can be generated by the system once for every block and contain information like the block number and other block parameters. This special event does not need to have an Inclusion and Subscription fee, and neither requires a signature. Smart contracts in the system may subscribe to this event, in order to be triggered automatically at certain block intervals.

\subsection{Gas Fee for Computation}
In a transaction-based system, the entity that generates a transaction has to pay for the gas fee associated with the computation, storage, and other costs of any smart contract code executed due to the transaction. This includes the contract code to which the transaction is sent and those contracts that the recipient code calls or invokes.

Such a design cannot be adapted with an event-driven model to avoid subscriber spamming.  Hence, the natural design is to have any subscriber pay for the gas fee associated with running its code. This is also in line with the event-driven paradigm’s space decoupling since the publisher does not have to concern itself with the subscribers to its event. The event publishers pay an inclusion fee (specified as part of the event data structure) each time they publish an event to the chain. This fee is specified explicitly if the event is published externally. In case the event is published internally i.e., through a smart contract, the fee is determined by the gas fee being paid by the publisher contract for its execution.

The other side of this problem is a malicious contract spamming the network with events and draining subscriber contracts of their ether. This can be addressed by allowing smart contracts to specify, at the time of subscription, to only run for events if generated by a particular public key (contain the corresponding digital signature) as described under heading ~\ref{sec:5.3}. Such a system works, for example, even if an oracle platform has multiple nodes because they can all still generate events with the same public key signature. Additional logic and using the block rate and event rate variables also help prevent these spams from occurring.

All smart contracts specify at the time of subscription the gas price they are willing to pay for their execution. The system prioritizes subscription execution, depending upon the gas price offered. Perhaps in the future, a provision can also be made for a floating gas price, with a maximum value and a weight value defined in the subscription parameters and the gas price computed by the network dynamically based on the variable system traffic and the constant weight parameter.

\subsection{Incentivization for Event Publishing}
Smart contracts that publish events have no incentive to do so unless they are being compensated. For example, an oracle interface contract providing external data to the system through events needs to be compensated for its services. In a transaction-based system, this is pretty straightforward. The user contract pays the interface contract when it interfaces with it i.e., generates the first transaction. In an event-driven system, this can be achieved by having the publisher describe a compensation rate each time it publishes an event. This is done through the subscription fee parameter in the event data structure. Each subscriber then has to pay the publisher the set fee in order for the system to execute the subscriber contract code’s subscription. So the subscriber pays both the miners and the publisher for the subscription execution. Subscribers specify the maximum rate they are willing to pay through the subscription fee field at the time of subscribing.

\begin{table*}[hbt!]
\footnotesize
\centering
\caption{Comparison of Transaction-driven and Event-driven models}
\label{tab:descomp}
\begin{tabular}{ p{2cm} | p{7.2cm} | p{7.5cm} }
\hline
\hline
                                       & \textbf{Transaction-driven model}                                                                                                                                     & \textbf{Event-driven model}                                                                                                                                                       \\ \hline
                           \hline
\textbf{Communication}                 & Smart contracts and external accounts communicate through transactions.                                                               & Smart contracts and external accounts communicate through events.                                                                                 \\ \hline
\textbf{Execution Trigger}             & Smart contract executions are triggered by transactions initiated through external accounts.                                                             & Smart contract executions are triggered by events generated by external accounts and smart contracts.                                       \\ \hline
\textbf{Computation Cost}              & Execution costs of the triggered contract and all its calls are paid by the transaction initiator. & Subscribing smart contracts pay for their own execution costs and specify the gas fee when subscribing.                                             \\ \hline
\textbf{Synchronization}               & All communication between participants is synchronous with the triggering transaction.                                                                                & All communication between participants is asynchronous.                                                                                                                           \\ \hline
\textbf{Token Transfer}                & Tokens are transferred between participants through transactions.                                                                                                     & Tokens are transferred between participants through a special transaction event.                                                                               \\ \hline
\textbf{Direct On-chain record}        & Only external account generated transactions are directly recorded on-chain.                                                                                          & Only external account generated transaction events are directly recorded on-chain.                                                                                                            \\ \hline
\textbf{Indirect On-chain record}      & The root of the transaction trie, state trie(includes storage trie) and receipts trie is referenced on chain.                                                         & The root of the transaction trie, state trie(includes storage trie), event state trie (includes event definition and event subscription trie) and receipts trie is referenced on chain. \\ \hline
\textbf{Inter-contract Sends} & Inter-contract token transfers are completed when triggering transaction is processed.                                                                      & All inter-contract token transfers are asynchronous.                                                                                                                              \\ \hline
\textbf{Transaction/Event Queueing}    & A pending transactions pool stores all the transactions that are yet to be processed.                                                                                 & A pending buffer stores all the events and subscription triggers that are yet to be processed.                                                 \\ \hline
\textbf{Transaction/Event Ordering}    & Miners are free to decide the order of the transactions in a block.                                                                                                   & Miners order the events/subscription triggers based on the gas price offered.                                                                                                     \\ \hline
\textbf{Transaction/Event effects}     & A transaction’s effects on the system state only occur when the transaction is processed.                                                                             & An event’s effects on the system state may occur indefinitely (until the queueing buffer clears).                                            \\ \hline \hline
\end{tabular}
\end{table*}

\subsection{Execution Independence and Atomicity}
In the event-driven paradigm, smart contracts only interact with each other through posting and listening to events. Smart contracts do not have to make calls and wait for the execution of other smart contracts before resuming their execution. This is the execution flow decoupling which the event-driven paradigm provides. Hence, once a smart contract starts executing in the proposed system, it completes its execution independently of other smart contracts. The system must ensure that the currently running smart contract finishes its execution before any new contracts triggered by the completed contract are run. The system guarantees atomicity in smart contract execution, and any exceptions raised in execution result in the current smart contract execution being reverted. However, no other contracts and their state is affected by the reversion.

This approach saves the systems from many troubles that exist in the transaction-based systems like cyclic executions and execution livelocks/deadlocks. This paradigm also prevents the publisher contract's execution state from being reverted by any subscriber contracts running out of gas or throwing an exception.

The system can still allow the reuse of smart contract code by having a similar opcode like that of DELEGATECALL in Ethereum. Since the called contract's code is executed in the state of the current contract, this will not violate the system's paradigm.

\subsection{Subscription Execution and Selection}
\label{sec:5.8}
In a transaction-based system, the selection of contracts to execute is relatively straightforward. Upon receiving a new block, any node begins executing transactions in the order that they are present in the block transactions list, and for each transaction execution, all subsequent in-lying smart contract calls and executions are executed first in the form of a LIFO stack.

The situation is somewhat trickier in an event-based system. If executing a smart contract generates one or more events, we have to decide which subscriptions to execute first. The previous heading has already established that a running smart contract execution will complete before any other subscriptions are processed. Hence in the event-driven system, a buffer is maintained of all pending subscription executions and events to process. We mandate that the list is ordered based on whichever subscriptions pay more gas fee. If two subscriptions pay the same gas fee, then the older defined subscription gets the preference. 

\begin{figure}
\begin{center}
\includegraphics[width=3.2in,angle=0]{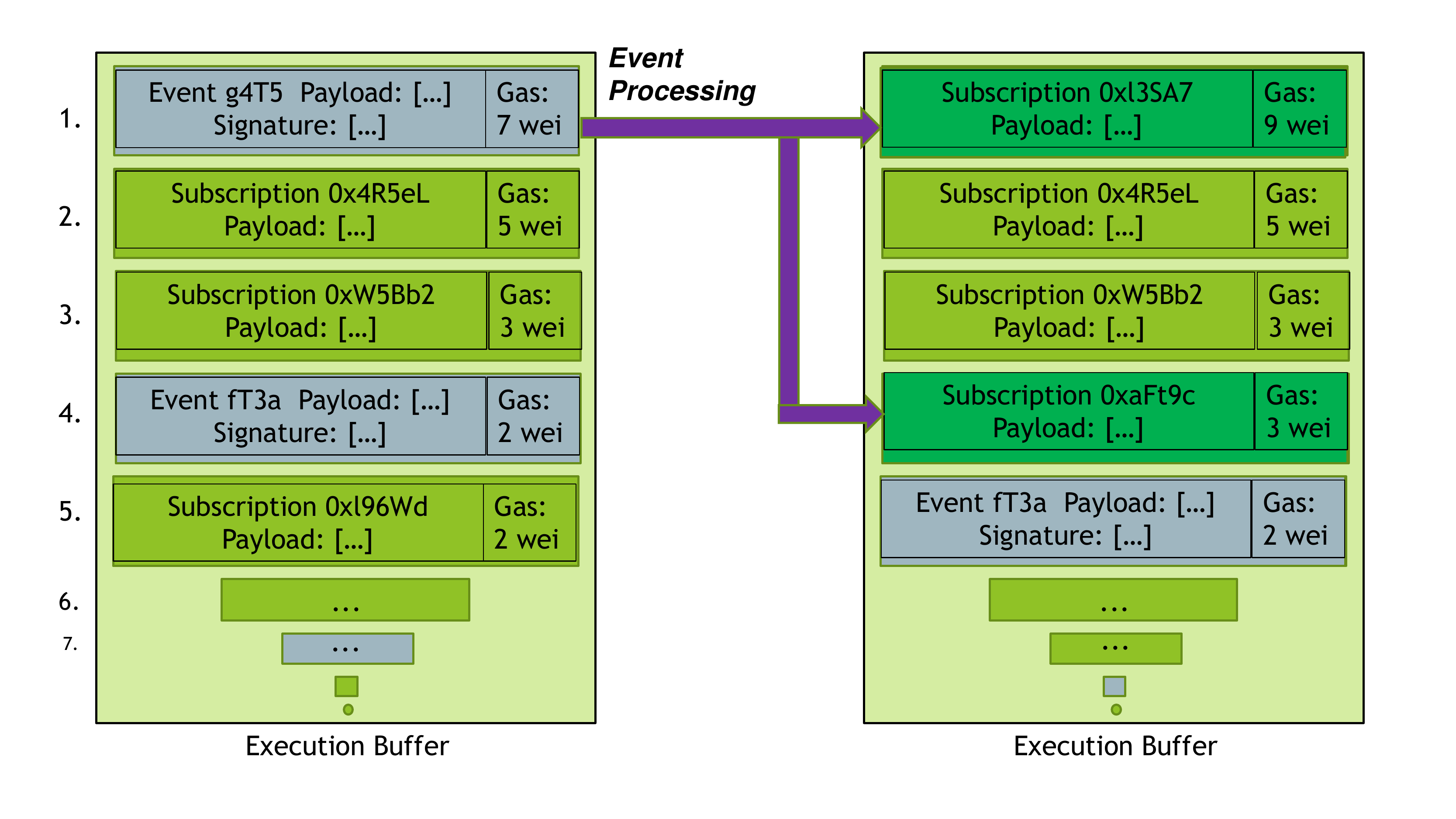}
\caption{Event processing in execution buffer.}
\label{fig:buffer}
\end{center}
\end{figure}

Whenever a new event is generated while running a smart contract, all its subsequent relevant subscriptions are added to the execution buffer, which is somewhat similar to the pending transaction pool in Ethereum. Subscriptions go into the buffer list, and their position is determined by the gas fee that they are paying. Whenever a subscription execution completes, the system will pick the next subscription from the list to execute. This buffer is not discarded between blocks and allows event-triggered subscription from previous blocks to run too, provided space is available. However, unlike the transaction-based model, there is no guarantee of a subscription being executed in the current or even succeeding blocks. Subscriptions paying more gas fee will always get the precedence in the system. Events that are generated externally have to be processed independently. The processing of an external event refers to determining the corresponding subscriptions to execute against it. External events also exist in the execution buffer competing for processing with subscription triggers, and the inclusion fee specified when generating these events determines when the system will process them. Once an event is processed by the system, its corresponding subscription triggers replace it in the pending executions buffer as illustrated in \figurename~\ref{fig:buffer}.

Having a deterministic fee-determined execution rule allows us to do away with putting all events on-chain. Since the system follows a gas price determined precedence rule for subscription executions, all nodes will arrive at the same state, and there is no need to put contract generated events on-chain. This allows the system to be as efficient in its chain space usage as the transaction-based system in the worst-case scenario. Any advantage provided by using event subscriptions is a bonus.

\vspace{-10pt}
 
\subsection{Block Validation}
For block validation, each node looks at only the events generated by the external accounts present in the event list for the current block. The node then proceeds to place these events in the pending subscription buffer. It then begins subscription executions/event processing from the pending subscription buffer, and subscription triggers for all the internal events generated by contracts are also placed in the buffer as they occur. The node keeps on executing subscriptions/event processing until the block gas limit is reached, or the buffer is empty. The node then compares the state, receipt, and event subscription/definition trie references to the ones provided in the block. If they match, the block is approved.

\vspace{-8pt}

\subsection{Parallel Processing and Sharding}
The system design allows it to be a better candidate for sharding and parallelization solutions than transaction-driven models. Since a smart contract being executed is not dependent on other contracts’ states, other contracts can be simultaneously executed in parallel on other shards. Because contract execution atomicity is guaranteed, there is no need for state locks. In case sharding is to be implemented, we propose having a shared subscription/event definition trie between the shards and a shared pending execution buffer. The shards can then divide contracts among themselves and only execute the relevant ones. Since the pending subscription execution list will be shared, they will have visibility to any events generated for the contracts in their domain. Any new events generated will be broadcast on the network for all shards to see.

A summarized comparison between the transaction-driven execution model and the event-driven design proposed in this section is presented in \tablename~\ref{tab:descomp}. 
\section{Security Considerations}
\label{sec:security}


\begin{table*}[hbt!]
\footnotesize
\centering
\caption{Summary of event-driven design security considerations.}
\label{tab:security}
\begin{tabular}{ p{0.1cm} | p{4.5cm} | p{9.0cm} }
\hline
\hline
\multicolumn{1}{c|}{\textbf{Layers}}                                     & \multicolumn{1}{c|}{\textbf{Threats}}                     & \multicolumn{1}{c}{\textbf{Analyses}}                                                                                                 \\ \hline
\hline
\multirow{4}{*}{\textbf{Economic}}                                       &  Market exploiting attack (from miners or event subscribers)      & \parbox{9cm}{Addressed by enforcing block delay for any update to event subscription states and Merkle hash.}                                 \\ \cline{2-3} 
                                                                         & Miner ordering attack to realize extractable values                               & Addressed by validation of scheduling of triggered smart contracts according to the global subscription states by peers (protected by hash of subscription states stored in block header).  \\ \hline
                                                                         
\multicolumn{1}{c|}{\multirow{5}{*}{\textbf{Programming \& Toolchain}}} & Reentrancy Attack                                        & \multirow{5}{*}{\parbox{9cm}{Addressed in the system by execution flow decoupling and having execution independence and atomicity.}}                 \\ \cline{2-2}
\multicolumn{1}{c|}{}                                                   & DoS with unexpected revert                               &                                                                                                                                        \\ \cline{2-2}
\multicolumn{1}{c|}{}                                                   & DoS with gas limit exceeded                              &                                                                                                                                        \\ \cline{2-2}
\multicolumn{1}{c|}{}                                                   & Unchecked call return value                              &                                                                                                                                        \\ \cline{2-2}
\multicolumn{1}{c|}{}                                                   & Call stack depth limit exceeded                          &                                                                                                                                        \\ \hline
\multirow{4}{*}{\textbf{Protocol}}                                       & Transaction ordering dependence                          & \parbox{9cm}{Addressed by having gas price based execution order for subscriptions and events.}                                 \\ \cline{2-3} 
                                                                         & Event publishing freeloading                             & Addressed by commitment scheme.                                                                                                      \\ \cline{2-3} 
                                                                         & Event spam attack on subscriber by publisher             & Addressed by allowing subscriber to specify subscription logic and frequency. \\ \cline{2-3} 
                                                                         & DoS by event spamming & Addressed by having gas fee associated with event registration and publication.  \\ \cline{2-3}                          & Fairness & Addressed by enforcing upper bounds of triggered smart contracts per user account and/or per event in each epoch.                                               \\ \hline
\textbf{Data}                                                            & Various                                                  & \multirow{3}{*}{Not event-driven design dependent, and addressed by blockchain protocols.}
\\ \cline{1-2}
\textbf{Consensus}                                                       & Various                                                  &                                                                                                                                        \\ \cline{1-2}
\textbf{Network}                                                         & Various                                                  &                                                                                                                                        \\ \hline
\hline
\end{tabular}
\end{table*}

The event-driven smart contract platform design offers numerous security benefits over a traditional transaction-driven model. 
For instance, when EDSC is integrated with Ethereum, the system can prevent inter-smart contract communication-related vulnerabilities, including reentrancy and denial-of-service attacks~\cite{chen2020survey}.
Since the proposed design provides execution flow decoupling, smart contract execution independence and atomicity, these vulnerabilities are no longer present in such a design. 
Aside from these vulnerabilities at the programming and toolchain layer, EDSC can also mitigate vulnerabilities arising from transaction ordering dependence~\cite{luu2016making}. 
EDSC achieves this by enforcing an order for event processing and subscription execution based on the gas fee. This design choice also provides the additional benefit of not having to put smart contract generated events on-chain and allows for reduced block confirmation times. Below, we discuss some attack scenarios and mitigation approaches. 


Denial-of-service (DoS) is a realistic risk for public blockchains. 
For instance, an attacker may try to flood the event processing module with event messages, or launch starvation attacks by polluting the event buffer.
These attacks can be mitigated with a variety of countermeasures, for instance, imposing a limit on event update and event creation rates for an account. 
In addition, event publishing also has a gas cost associated with it, which is paid by the publishers, to discourage them from publishing events unnecessarily.  Depending on the gas limit of an event, a registered event can be kept in the system for only a bounded number of blocks (limit can be re-fueled later by the creator with a transaction). Similarly, generating a new subscription/event definition and deploying a new contract also have an associated gas fee, which the event publisher must pay analogous to Ethereum’s associated gas fee for deploying a new contract. 
These fees serve as a deterrent against spamming and DoS attacks on the system. Furthermore, event manager enforces that for each event and user account, there is a maximum number of transactions that can be triggered in each epoch, which prevents event buffer pollution. 
To further mitigate the risk of event publishers spamming the system, EDSC allows subscribers to use variables like the Event Rate and Block Rate as well as the Subscription Logic expression to control their frequency of subscription execution.

Malicious market exploiting and related cheating behaviors are another type of threats.
In many DeFi applications such as DEX, a smart contract is applied to execute financial transactions. Such systems are exposed to various market-exploiting behaviors (e.g., frontrunning)~\cite{daian2019flash}. 
Similar market-exploiting behaviors may pose a risk to EDSC. For instance, when a node observes an event update where financial value can be extracted, the node may send a shortcut message that registers to the event or updates its event registration to boost its priority in the event buffer. Similarly, when a miner detects an opportunity that value is extractable, the miner may be incentivized to directly insert a new event subscription or modify existing subscriptions. Since miner controls event processing, the miner may take advantage of this position to order events or transactions generated from event subscriptions in a particular epoch in ways to extract values besides block reward and transaction fees. In EDSC framework, such attacks are prevented by the global event subscription state. The event subscription state is protected with Merkle hash tree and the root hash is included as part of  block header. Updates to the global event subscription state are initiated through on-chain transactions and confirmed using the underlying blockchain consensus mechanism. The EDSC system enforces minimal delay for changes to the global event subscription state to be effective (minimal block delay). When a block is propagated and received by a peer, the peer will validate the transactions that are triggered by events according to the global event subscription state. This means that any foul play or dishonest manipulation of event triggered transactions can be detected by other peers and the block will be rejected. 

EDSC is also susceptible to ``freeloading'' risk,  i.e.,  freeloaders in the system can observe the events being published and copy the payload and publish the same events themselves at a lower subscription fee. 
This problem also exists for oracle systems like ChainLink~\cite{chainlink}.
Several methods can be applied to address this issue.
For instance, ChainLink uses a commitment scheme to prevent such attack, which can be easily incorporated into EDSC.

Since EDSC can be implemented on any smart contract platform, vulnerabilities that are present in smart contract itself are not considered.
We summarize all the security analysis in \tablename~\ref{tab:security}.
\section{Implementation}

\subsection{Event Enabled Blockchain Node}

Design of EDSC model can be implemented by extending Ethereum blockchain. We used Golang implementation of Ethereum client as the target. Extensions include new messages for creating events, sending event updates, managing event subscriptions, processing events, generating new transactions in real-time based on event updates, and etc. 

In Ethereum, P2P module is responsible for communicating with the underlying P2P network using a gossip-type strategy.  It receives and routes various messages through communication with the neighbors (nodes that are peers) under protocol manager. In case of EDSC, in addition to blocks and transactions, a node receives and delivers event-related messages and transactions to the extended protocol manager module that handles event-related messages and decides the next step of processing. There are dedicated messages for event creation, event updates, event subscription, event unsubscription, and event subscription updates. Except for event updates, event messages are special transactions that affect the global event states. A gas fee is charged for operations such as creating an event, making an event subscription, or updating an existing event subscription. 

There is a dedicated address for receiving event messages. Event messages are signed using ECDSA and secp256k1 digital signatures by the senders.  There is a nonce in each event-related message. Event creation transaction is used to register an event, identified with a 160-bit long unique address (created from the sender’s account address and event definition).  In addition to event payload data, each event update message identifies its associated event address and sender’s address. 

Each node implements an event buffer and event manager for processing event messages. When a node receives a new event message, its ProtocolManager module first sends the new message to the event manager for validation, including verifying signatures and checking other constraints and security requirements such as event update rates. When validation is passed, event transactions will be forwarded to the PendingPool of Ethereum TxPool where incoming and pending transactions are stored. Event update messages are handled by the event buffer (evtBuffer) because event updates are not processed as transactions. 

TxPool and event manager notify the ProtocolManager module that there is a new event message that can be forwarded to other neighbors. Then, the ProtocolManager randomly selects $\sqrt{N}$ downstream peers that do not know the event messages as the targets to forward this message. For the remaining N - $\sqrt{N}$ downstream peers, the event message hash will be forwarded. A peer will receive event message hashes from its neighbors. When a node randomly selects one of the neighbors that have sent it the new event message each peer receives the hash of a new message, the node waits for a while (e.g., 500 ms). During this period, if there is no other neighbor sending the same event message to it, the nsh, and sends a GetEvt information to the selected neighbor for requesting the new event message. After the requested neighbor returns the event message, the node first validates it.  After validation, it is either added to the TxPool or evtBuffer of the node. 

\begin{figure}
    \begin{center}
    \includegraphics[width=3.5in,height=2.4in]{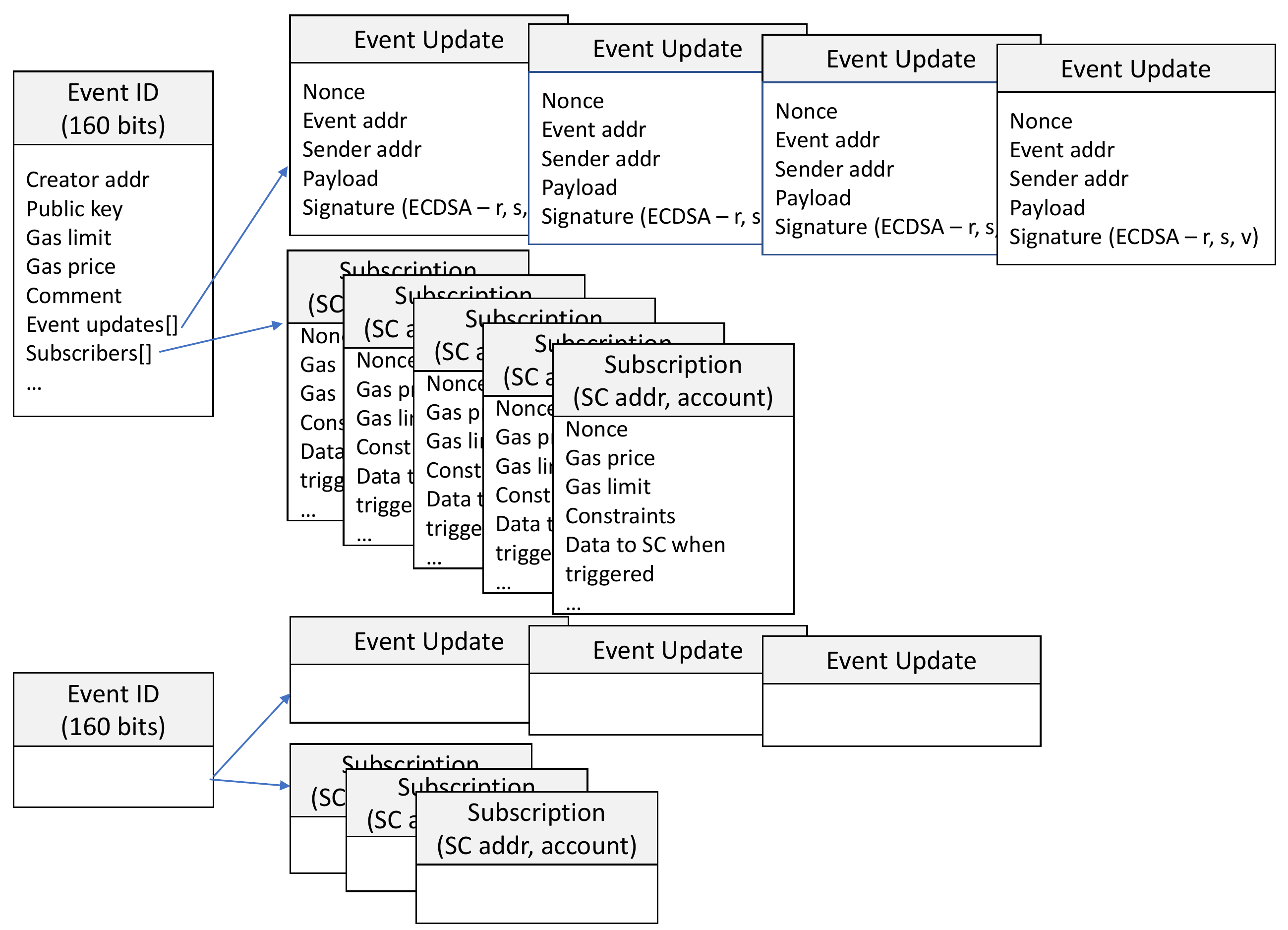}
    \vspace{-1em}
    \caption{Illustration of event data structures and states. }
    \label{fig:sub}
    \end{center}
    \vspace{-1em}
\end{figure}

The extended protocol manager module processes the received event messages and delivers event updates to evtBuffer. A node maintains and keeps track of event subscription state, as illustrated in \figurename ~\ref{fig:sub}. There is a map where the node can retrieve a list of subscriptions for each event update. Each subscription links to a smart contract and user account. For each event address, subscriptions are ranked based on priority (e.g., using gas price or other metrics of priority). For each event address, event updates are ordered using nonce. For each epoch, based on the event subscription state and events in evtBuffer, a new set of event triggered transactions are created.  The event triggered transactions are added to the node’s PendingPool. In Ethereum client, PendingPool maintains the pending transactions that have not been included in the blocks on the blockchain but are ready to be packaged into a new block. Similar to how PendingPool tracks pending transactions for each account,  the set of event triggered transactions enforces separate upper bounds for number of transactions per event update and number of transactions for account. 

\begin{algorithm}
  \small
  \SetKwInOut{Input}{Input}
  \SetKwInOut{Output}{Output}
  \Input{tx pool $\mathit{txPool}$,  event updates $\mathit{newEvts}$, event subscription state $\mathit{evtSubState}$} 
  \Output{$\mathit{block}$, updated $\mathit{txPool}$}

	{\it Create a new empty $\mathit{block}$} \\

	{\it Set $\mathit{tmpEvts}$ = $\mathit{newEvts}$} \\
	
	{\it While } \\

	{\it \hspace*{5mm} stop when $\mathit{txPool}$ is empty   }\\
	
	{\it \hspace*{5mm} stop when $\mathit{block}$ limit is reached (gas limit or block size) }\\
	
    {\it Set $\mathit{tmpEvts}$ = validate-and-filter-evts($\mathit{tmpEvts}$) }\\
    
    {\it Set $\mathit{newTxs}$ = create-tx-based-on-evts($\mathit{evtSubState}$, $\mathit{tmpEvts}$)} \\
		 
	{\it Set $\mathit{txPool}$ = merge $\mathit{newTxs}$ with $\mathit{txPool}$}\\
	
    {\it Set $\mathit{pendingTxs}$ = tx-filter($\mathit{txPool}$) } \\
    
    {\it Set $\mathit{sortedTxs}$ = sort($\mathit{pendingTxs}$)  } \\

    {\it Set $\mathit{selectedTxs}$ = pick top n best $\mathit{txs}$ from $\mathit{sortedTxs}$ } \\

    {\it $\mathit{block}$, $\mathit{tmpEvts}$ = execute-txs($\mathit{selectedTxs}$, $\mathit{block}$)} \\
	
	{\it End while loop }\\

	\caption{Event processing algorithm.}
	\label{alg:event}
\end{algorithm}

After event triggered transactions are added to the PendingPool, the way that they are sorted and picked for execution and block creation very much follows the same design of Ethereum client. In case of PoW, the incentive includes block reward and transactions fee. A miner uses the gas mechanism to calculate the fee for transactions with smart contracts. To determine the fee for transactions and blocks, it uses attributes such as gas limit and gas price. In short, used gas multiplied by the gas price corresponds to the fee that the miner receives, where used gas depends on the computational requirements of the smart contract ~\cite{baird2019economics,8530775}, but never exceeds the gas limit.

\begin{table}
\centering
\caption{Operations}
\label{tab:op}
\scriptsize
\begin{tabular}{p{0.25\linewidth} | p{0.6\linewidth}}
\hline
\hline
 \textbf{Operation}                              & \textbf{Meaning}   \\ 
\hline
\hline
validate-and-filter-evt &   Validate and filter events including verification of signatures and constraints such as rate of event updates.       \\
\hline
create-tx-based-on-evts & Create txs based on event subscription states (enforce rules such as k txs at most for each event based on priority).        \\
\hline
tx-filter & Filter pending txs, for instance m txs at most for each account. \\
\hline
sort  & Sort txs based on priority (e.g., gas fees). \\
\hline
execute-txs  & Execute pending txs and add to block.  \\
\hline
\hline
\end{tabular}
\end{table}

\subsection{Modeling Tools}

For modeling EDSC and experimenting with design options in scalable manner, we extended BlockSim ~\cite{Alharby_2020}, a framework and software tool based on discrete-event dynamic models for blockchain systems. BlockSim supports the analysis of a variety of blockchain deployments as well as for design exploration and experimentation. It implements models for Bitcoin, Ethereum and other consensus algorithms. Results of BlockSim have been validated by comparing with design properties and measurement studies available from real-life blockchains such as Bitcoin and Ethereum. We modified BlockSim’s full modeling technique for Ethereum to support the EDSC framework and event triggered transactions. The model includes events, event subscriptions, transactions, blocks, transaction pool, and blockchain ledger.  Events and transactions created by a node are propagated to all other nodes in the network. Upon receiving the event or transaction, the recipient node appends it to the corresponding pool/buffer for event or transaction processing. 

\section{Experiment Results and Analysis}

We conducted experiments with the extended clients and BlockSim modeling tool. The implemented model of BlockSim for Ethereum has been validated using real data ~\cite{Alharby_2020}. The model takes a set of parameters as inputs. These current implementation of Ethereum baseline model compromises of 12.42s block interval and 2.3s block delay ~\cite{Alharby_2020}. The model is configured to use the same parameters as currently in Ethereum.  The results are based on averages of independent simulation runs of 10,000 blocks. We compare EDSC smart contract delay with the baseline delay of oracle contract based design. The delay is measured as the time when event update is sent by an oracle node to the time that the transaction of a triggered smart contract is added to a block of the longest global chain. 

As indicated by the results, EDSC achieves shorter delay for running contracts that subscribe to events, on average often less than time of block interval. In contrast, the baseline model incurs delay longer than three blocks (similar delays observed in Ethereum contracts in real life using oracle contracts: +3 block delays - see \figurename ~\ref{fig:OracleGraph}). This pattern is observed under different block intervals, varying from 8s to 60s. 

\begin{figure}
    \begin{center}
    \includegraphics[height=1.2in,width=2.8in]{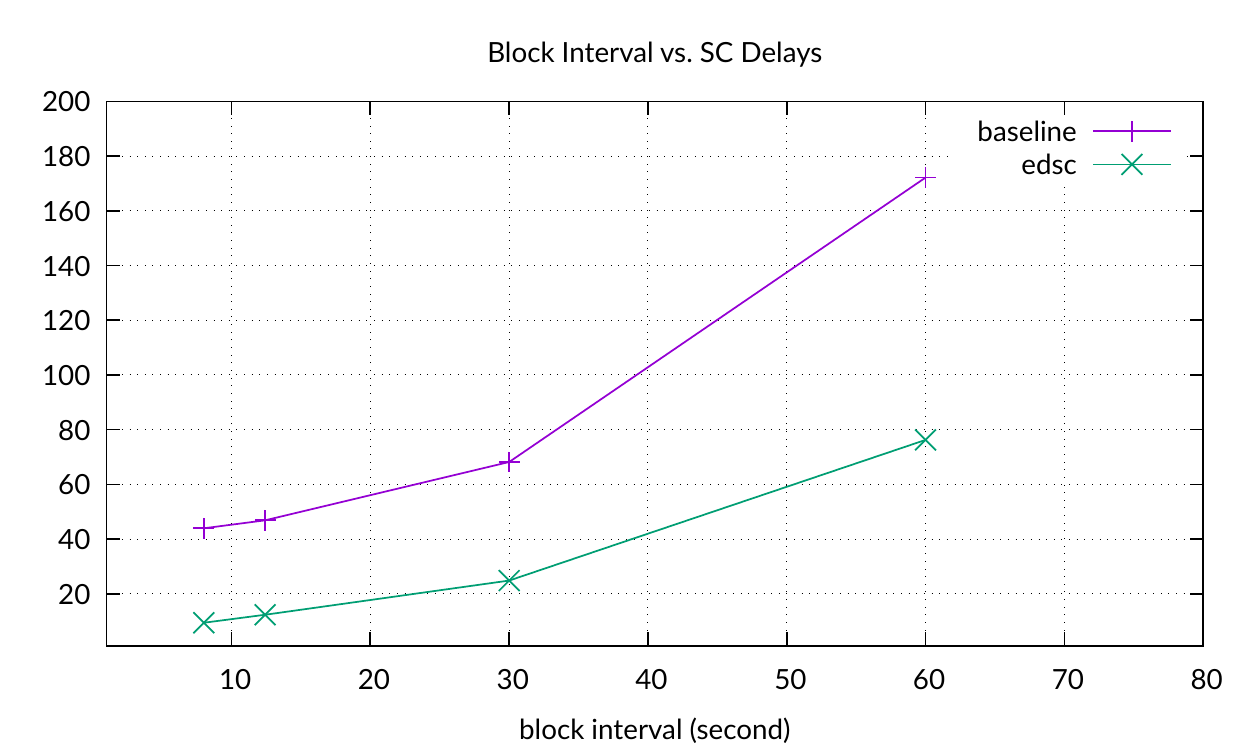}
    \vspace{-1em}
    \caption{Block intervals and SC delays (seconds). }
    \label{fig:binterval}
    \end{center}
    \vspace{-1em}
\end{figure}

\begin{figure}
    \begin{center}
    \includegraphics[height=1.2in,width=2.8in]{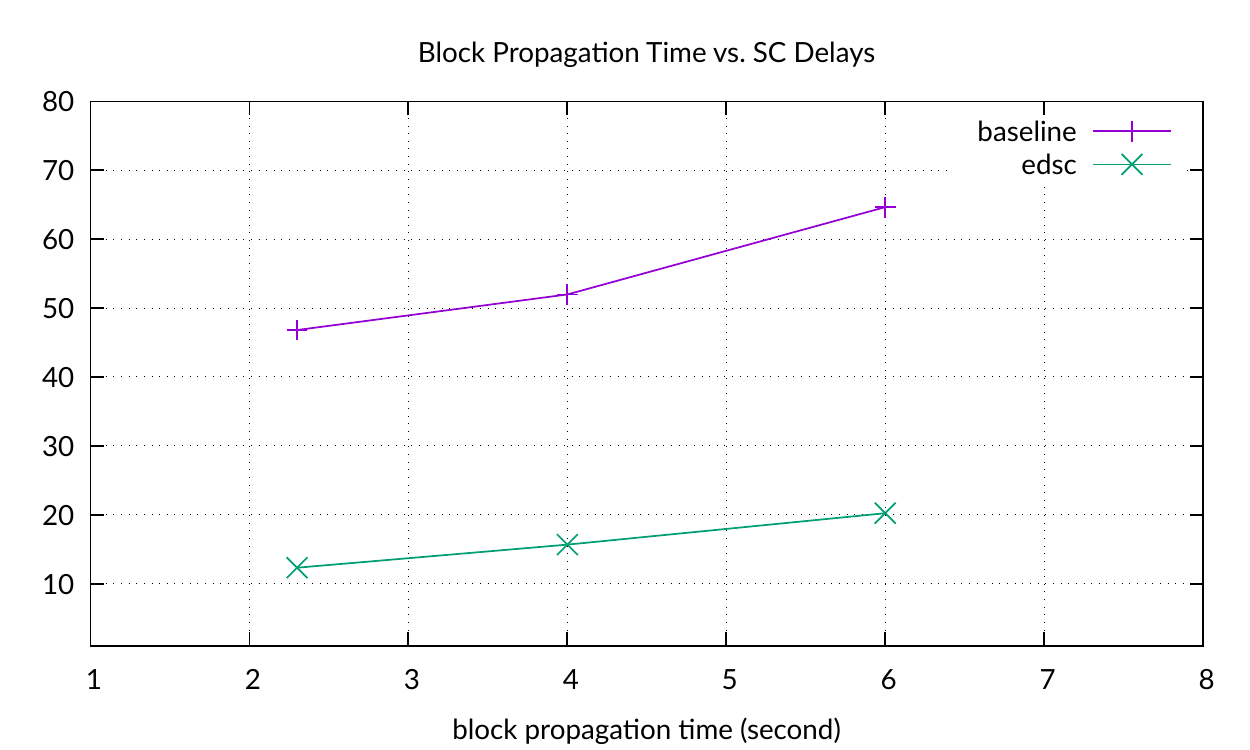}
    \vspace{-1em}
    \caption{Block propagation time and SC delays (seconds). }
    \label{fig:bdelay}
    \end{center}
    \vspace{-1em}
\end{figure}

\begin{figure}
    \begin{center}
    \includegraphics[height=1.2in,width=2.8in]{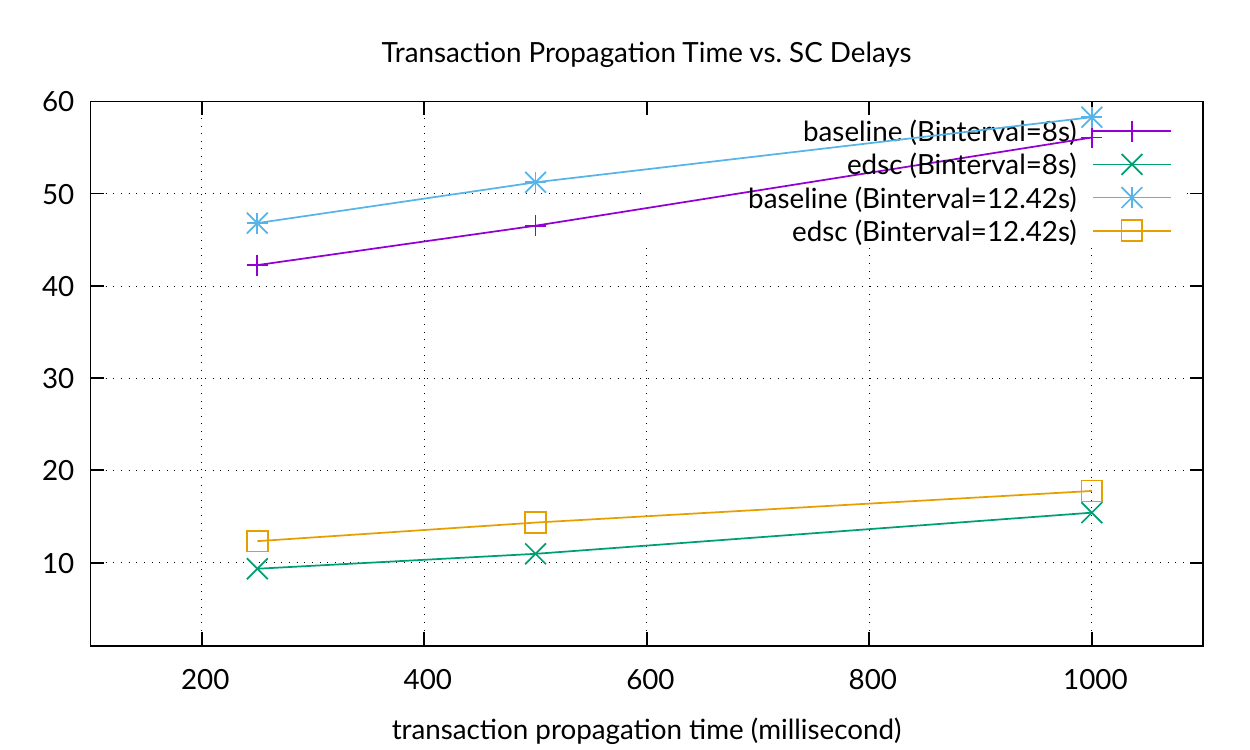}
    \vspace{-1em}
    \caption{Event/transaction propagation time (millisecond) and SC delays (seconds). }
    \label{fig:tdelay}
    \end{center}
    \vspace{-1em}
\end{figure}

Both block delay and event/transaction delay can affect the latency between the event update and inclusion of transactions from the triggered contracts. One can assume that this latency likely increases when either block delay or event/transaction delay grows larger. Results in \figurename \ref{fig:bdelay} and \ref{fig:tdelay} confirm this hypothesis. However, delays in the baseline model appear to be more affected negatively by block delay or transaction delay increase as illustrated by the expanding distance between EDSC delay and the baseline model delay.

\begin{figure}
    \begin{center}
    \includegraphics[height=1in,width=2.6in]{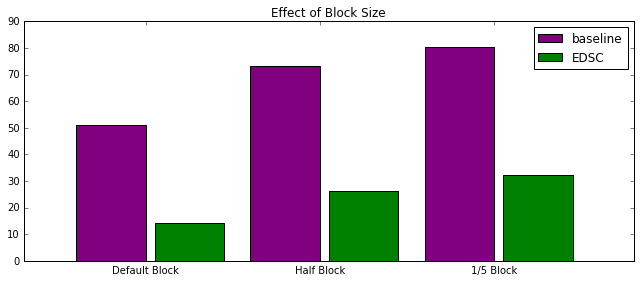}
    \vspace{-1em}
    \caption{Block capacity and SC delays (seconds). }
    \label{fig:bsize}
    \end{center}
    \vspace{-1em}
\end{figure}

Other factors may also have influences on latency of event driven contracts. Block size is one such factor. As suggested by results in \figurename\ref{fig:bsize}, decreasing block size will negatively impact contract latency under both models. However, the latency benefit of the EDSC model over the baseline is not affected. 

On average, delays under the EDSC model could be from 2.2 to 4.6 times less than the delays of the baseline model (depending on block interval, block delay, etc), which demonstrates its effectiveness for supporting contracts that demand timely execution based on events.

\section{Example Use Cases}
\label{sec:exampleusecases}

The event-driven paradigm is, by design, a better fit for many emerging smart contract applications. The event-triggered execution, asynchronous communication, and contract execution independence and atomicity features are particularly instrumental in building scalable, adaptable, and easy to maintain applications on the smart contract enabled blockchains. The paradigm also enables these applications to be more reactive to external or internal triggers without overloading the system. In particular, financial applications like algorithmic trading, deploying financial instruments, real-time analysis, or digital asset management are naturally suited for the event-based model. The model also has been proven instrumental in a diverse application range consisting of supply chain management, online betting, oracle systems, gaming, etc. ~\cite{hinze2009event}. Here we elaborate on the design's benefits by discussing two broad real-world smart contract use cases.

\paragraph{Digital Asset Trading and DeFi Applications} DeFi applications relay on third parties to report real-time information about the market price of the assets from real-world (off-chain) sources ~\cite{liu2020look}. Consider implementing a digital asset trading platform on a blockchain-based smart contract platform. The system would need regular and timely updates on various market indicators like stock prices, trade volume, market trends, etc. Most of this external data is retrieved by employing oracle systems. In a traditional transaction-based system, this would require tedious and meticulous interfacing with multiple oracle system interfaces. Regularly managing the application would also not be easy. The implementer would have to figure out the interfacing details multiple times and familiarize themselves with the data formatting across multiple interfaces and providers. The two-way transactions for oracle fetches would burden the system if such applications were widely deployed.

In an event-driven platform, the integration is much simpler, cleaner, and easier to manage. The subscriber needs to know only the trusted publisher's address and the identifier of the event that they are interested in. The publisher might be a single entity or an oracle system. The event payload format and documentation would be available on-chain and would not differ if multiple sources were publishing the same event. For example, if ten smart contracts are listening for a particular stock price from a publisher, it would not result in ten or twenty transactions going on-chain. Instead, only one external event is recorded on-chain, and all ten contracts can run by subscribing to this one event. In addition to a cleaner interfacing mechanism, easier maintenance, and lesser data on-chain, the event-driven paradigm also allows subscribers to listen for particular transactions. Since transactions are just a type of event in our design, participants using such financial applications might subscribe to transactions only from a particular party, only to a particular party, or random transactions exceeding a particular amount, etc. This is not achievable in the transaction-driven model. A transaction model might use external listeners to observe such events on the chain and then make transactions to trigger specific executions but cannot do it without incurring a block delay. We looked at the transaction data for ChainLink~\cite{chainlink}, which is the most popular oracle service provider for DeFi applications and see it having a 3-4 block delay on average while responding to oracle requests in the last eighteen months as shown in \figurename ~\ref{fig:OracleGraph}.
\iftrue
\begin{figure}
    \begin{center}
    \includegraphics[width=3.0in,height=1.5in]{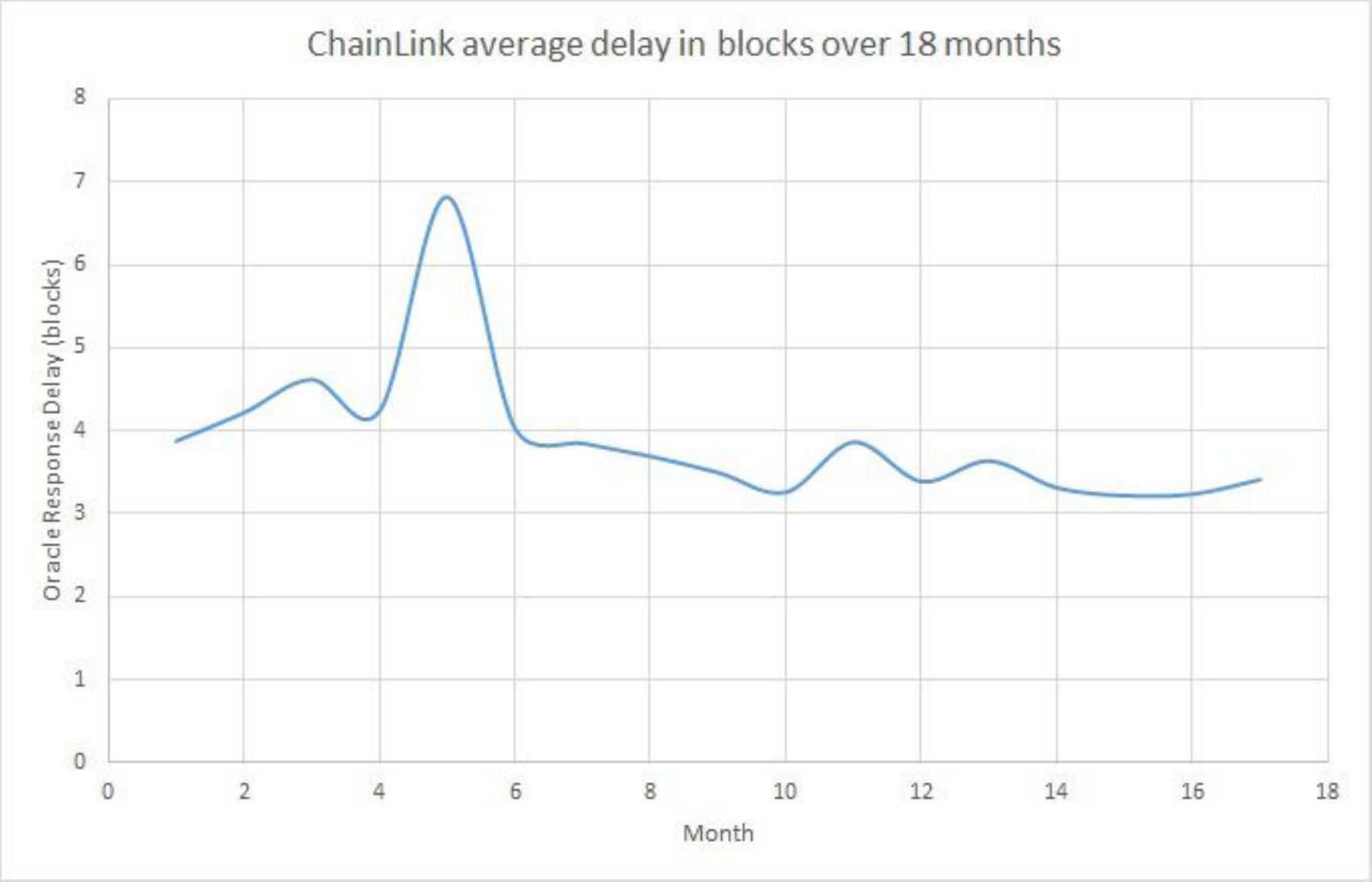}
    \vspace{-1em}
    \caption{ChainLink response average block delay (May 2019 to Oct 2020).}
    \label{fig:OracleGraph}
    \end{center}
    \vspace{-1em}
\end{figure}
\fi

\paragraph{Prediction Market Application} Similar to the first scenario, a prediction application on the blockchain also benefits from an event-driven paradigm's features. Smart contracts can lay dormant unless executed by external events like the result of a sports match or an election. For long-term or small bets, it might not be feasible to use the transaction model to poll for these external events or pay the fee for interfacing with an oracle system. The event-driven design makes such applications more feasible for smaller amounts since the event generation cost (maybe from a reputable news agency) is spread over numerous participants (subscribers). Again, the model also allows such an application to, for example, monitor newly placed predictions and adjust odds accordingly.

Hence, it shows that for many use cases, the event-driven design would be more cost-efficient (both computation and oracle fee), scalable, cleaner to implement, easier to maintain and allow for applications to have greater visibility on-chain data and token exchange.

It is worth mentioning that research on oracle service is complementary to event driven model of smart contract execution. These two are related but separate research topics. Our system can integrate various types of oracle services such as TEE based oracle service ~\cite{cryptoeprint:2016:168}, oracle service employing secure multi-party computation ~\cite{zhang2019deco}, decentralized oracle service, etc. In fact, an event driven smart contract platform can arguably provide better support for integrating oracle services.

\section{Conclusion} 
\label{sec:conclusion}
We proposed the concept for a novel event-driven smart contract platform with built-in event processing support on the blockchain. We presented a basic design as well as implementation of such a system in practice and commented on its potential benefits to smart contract use cases. We also presented analysis on its security aspects. Experiment results based on BlockSim extension are shown to illustrate performance advantages of event driven smart contract model. Being the first attempt to combine the two avenues of blockchain-based smart contracts and event-driven design, our work paves the way for future research on the subject in various directions. Future work can explore the application of this paradigm to implementing a sharding solution for scalability.


\bibliographystyle{IEEEtran}
\bibliography{references} 

\end{document}